# Validation of Compact Models of Microcantilever Actuators for RF-MEMS Application

Eugenio Brusa[1], Antonio Della Gaspera, Mircea Gh. Munteanu[2],
Dept. Elect., Manag. Mech. Engineering - Università degli Studi di Udine, via delle Scienze 208 – 33100 Udine, Italy;
[1]eugenio.brusa@uniud.it, [2]munteanu@uniud.it, phone [1]++39-0432-558299, [2]++39-0432-558243, fax ++39-0432-558251

*Abstract*-Microcantilever specimens for in-plane and out-of-plane bending tests are here analyzed. Experimental validation of 2D and 3D numerical models is performed. Main features of in-plane and out-of-plane layouts are then discussed. Effectiveness of plane models to predict pull-in in presence of geometric nonlinearity due to a large tip displacement and initial curvature of microbeam is investigated. The paper is aimed to discuss the capability of 2D models to be used as compact tools to substitute some model order reduction techniques, which appear unsuitable in presence of both electromechanical and geometric nonlinearities.

## I. INTRODUCTION

Reconfigurable layout is a typical feature of circuits used for radio-frequency application [1]. This goal is achieved by means of cantilever and double clamped microbeams, electrostatically actuated, being used as switches, resonators and varactors [1-4]. Design needs for a precise prediction of pull-in condition and frequency response. An effective modelling of these static and dynamic behaviours is rather difficult, because of the electromechanical coupling. Electromechanical forces nonlinearly depend on mechanical displacement, electric charge and voltage. In presence of large strain or displacement a structural nonlinear solution has to be implemented [5-7]. Analytical solutions for microcantilever and double clamped microbeams were formulated and included the effects of stretching and large displacement, i.e. the so-called geometrical nonlinearity [5,6]. Corrections suitable to predict fringing effect of electric field were even proposed [1,3]. These models assume an ideal beam geometry, which differs from the actual structure in some details of the constraint and electrode shape. Microcantilevers currently proposed by industry look like specimens depicted in Figs.1 and 2. A first geometry is based on "in-plane" bending test, i.e. microbeam deflection occurs in a plane parallel to the profiling system's target. In "out-of-plane" bending actuation microbeam tip moves towards the target [8]. Often microbeam is a part of a wider structure, e.g. it connects the rigid plate of a varactor to the fixed frame [9]. All these aspects motivate the need for a library of structural numerical models within the simulator used to predict the response of the whole electronic circuit. Accuracy and fastness are main requirements for this modelling activity. Authors demonstrated in previous papers that static pull-in of in-plane bending specimens is accurately predicted by a numerical solution based on sequential approach with voltage increments [7,10,11]. Geometric nonlinearity becomes relevant for large displacements, close to pull-in voltage. A double nonlinear solution, including geometrical effect, has to be implemented. In this case iterative solution is applied at each step of the sequential approach and makes extremely heavy the computational effort. Iteration can be avoided by using a special finite beam element and a non-incremental formulation [10,12,13]. Effectiveness of 2D FEM models in case of in-plane bending specimens was a little bit surprising, because no electric force concentration due to the finite dimensions of the geometry was taken into account [11]. In this paper a detailed analysis of in-plane microcantilevers is performed to complete that investigation. Results are compared to those of out-of-plane microcantilevers, which show an initial curvature of microbeam. Main features of the two above mentioned layouts are described. To define suitable criteria to proceed with a model order reduction useful for the dynamic analysis, limits of the coupled field analysis in 2D models are discussed, by investigating some three dimensional effects of the electric field. Numerical results are simultaneously compared to the experiments done on microspecimens made of epitaxial polysilicon and gold.

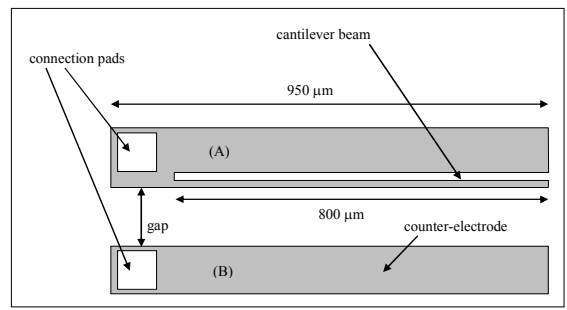

Fig. 1. Microcantilever built for in-plane bending test (top view).

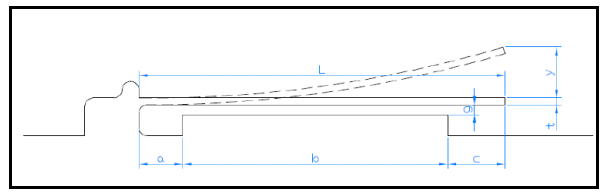

Fig. 2. Microcantilever built for out-of-plane bending test (front view).

  

## II. SPECIMENS FOR IN-PLANE BENDING TEST

Experimental validation of numerical models developed in previous papers was performed on eight geometries, described in Table I. FEM static analysis was performed in ANSYS code, as sequential solution with mesh morphing in dielectric region, then through the non-incremental FEM sequential approach tested in [10,12,13] and by a combined sequential FEM/BEM solution [11]. Last two methods were implemented in Matlab. An additional comparison included results of the Discrete Geometric Approach (DGA), recently proposed [14]. As Fig.3 shows in case of geometry 4, all the 2D approaches converge to the experimental curve, provided that Young modulus of epitaxial polysilicon was clearly identified and geometrical nonlinearity due to the large tip displacement was included. FEM 3D (ANSYS, SOLID122 electrostatic, SOLID185, elastic) analysis revealed that the actual distribution of the electrostatic force is fairly different in terms of average and peak values (Fig.4). Nevertheless, effectiveness of 2D models appeared surprisingly good. A deeper investigation of three dimensional effects of electric field allowed finding that agreement between 2D models and experiments was assured by constraint and electrode geometries.

TABLE I
MICROCANTILEVER SPECIMENS FOR IN-PLANE BENDING TEST

| n. | Length [μm] | Width [μm] | Thickness [μm] | Gap [μm] |
|---|---|---|---|---|
| 1 | 101 ± 0.1 | 15 | 1.80 ± 0.02 | 5.0 ± 0.3 |
| 2 | 101 ± 0.1 | 15 | 1.80 ± 0.02 | 10.0 ± 0.3 |
| 3 | 101 ± 0.1 | 15 | 1.80 ± 0.02 | 20.1 ± 0.3 |
| 4 | 205 ± 0.2 | 15 | 1.90 ± 0.02 | 10.0 ± 0.3 |
| 5 | 205 ± 0.2 | 15 | 1.90 ± 0.02 | 20.0 ± 0.3 |
| 6 | 805 ± 0.5 | 15 | 2.70 ± 0.04 | 39.6 ± 0.3 |
| 7 | 805 ± 0.5 | 15 | 2.70 ± 0.04 | 200 ± 0.5 |
| 8 | 805 ± 0.5 | 15 | 2.70 ± 0.04 | 400 ± 0.5 |

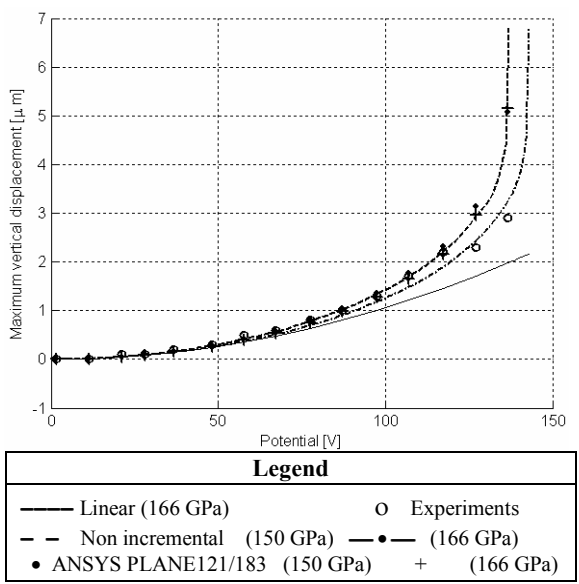

Fig. 3. Example of experimental validation on in-plane microcantilever

Legend:
— Linear (166 GPa)      o Experiments
– – Non incremental (150 GPa)   —•— (166 GPa)
• ANSYS PLANE121/183 (150 GPa)   + (166 GPa)

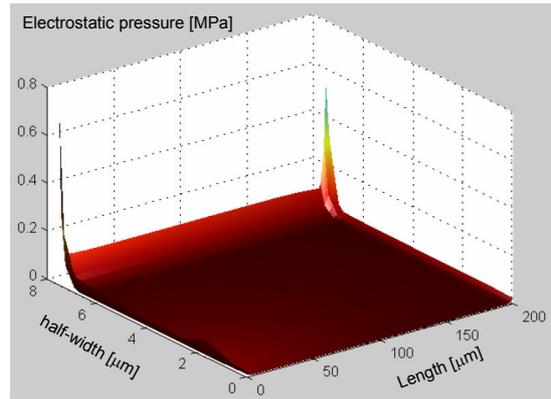

Fig. 4. Actual distribution of electrostatic pressure on half-width of in-plane microcantilever according to a 3D FEM model.

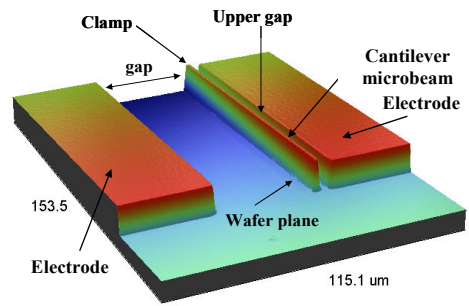

Fig. 5. Crucial aspects of in-plane microcantilever

In case of geometry 5, where half-width is $c = 7.5$ μm, ratio between the peak values of force computed by 3D FEM model and 2D respectively was 1.88. This result assumed that actuation voltage was applied to the counter-electrode, equal width for electrode and counter-electrode, rounded tip, no surface behind the beam. All these assumptions play a significant role in case of in-plane actuators (Fig.5). Charge concentration at the tip and along the edges increases electrostatic force, more largely as peak than as average value. This concentration is localized on a small area (Fig.4). Rectangular and rounded tips have higher force ratio 3D/2D than sharp triangular tip. Indefinitely long counter-electrode and zero voltage applied on the microbeam bring the above force ratio up to 2.1. Actually, wafer surface below the lateral edge of microcantilever decreases this ratio to 1.05, thus allowing 2D model predicting the actual pull-in. In case of equal width for upper and counter-electrode ratio tends to 1, being higher when counter-electrode width is larger than the electrode's one.

Geometric characterization of in-plane microcantilever is rather difficult. Profiling system offers a very high resolution along the optical axis, while it is lower on the target plane. Thickness and gap measurements for in-plane microcantilever is less accurate, as Table I shows. Numerical prediction of pull-in is consequently ineffective, if nominal values of these parameters are inputted. For a given correspondence of numerical and experimental values, and in case of geometry 5 following discretizations were set used: DGA (FORTRAN, sequential method) 15000 elements, 32000 DoFs (electrical);

756 elements and 4000 Dofs (structural); FEM (ANSYS; iterative, mesh morphing) 80 PLANE183 (solid beams), 3000 PLANE 121 (electrical); FEM (MATLAB; sequential; non incremental; mesh morphing) 3036 elements, 346 nodes (electrical); 41 nodes, 40 Timoshenko two-node beam elements; FEM/BEM (MATLAB, sequential, non-incremental) 337 two-node boundary elements, 188 nodes (electrical); 31 nodes, 30 two-node Timoshenko beam. As it is clearly described in [15], BEM allows reducing the number of DoFs in dielectric region, but boundary elements assure a better accuracy within the element field than on boundaries. To predict accurately voltages on the microbeam a mesh refinement is required, although DoFs are less than in FEM.

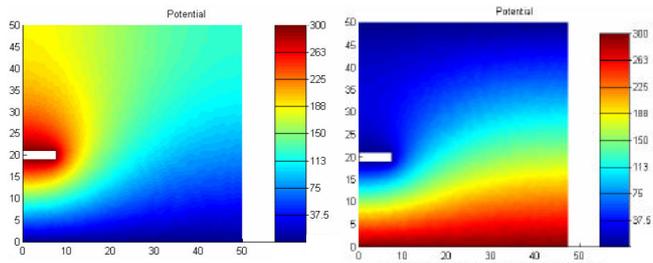

Fig. 6. Potential distribution around the microcantilever half-width in case of null voltage applied to counter-electrode (left) or to beam surface (right).

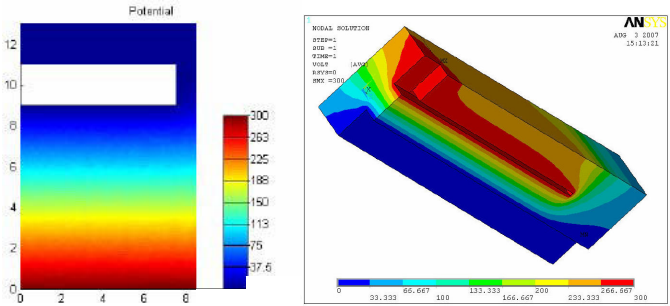

Fig. 7. 2D and 3D predictions of potential distribution around the microcantilever half-width in case of null voltage applied to beam surface and in presence of wafer surface (right edge in 2D model).

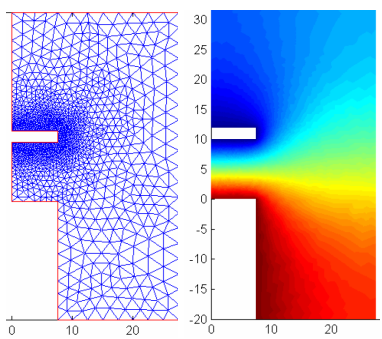

Fig. 8. Effect of finite dimensions of counter-electrode on in-plane microcantilever half-width.

### III. SPECIMENS FOR OUT-OF-PLANE BENDING TEST

Four geometries of golden microcantilevers for out-of-plane bending were built to perform a parametric analysis. Each one included several specimens. Table II summarizes relevant parameters. Numerical data are written by describing the range of measured values among different specimens and the measurement errors.

TABLE II
MICROCANTILEVER SPECIMENS FOR OUT-OF-PLANE BENDING TEST

| n. | Length [μm] | Width [μm] | Thickness [μm] | Gap [μm] | y [μm] |
|---|---|---|---|---|---|
| 9 | 531:535 ± 0.3 | 32:33 ± 0.3 | 2.9:3.0 ± 0.5·10⁻⁴ | 2.88:2.99 ± 0.5·10⁻⁴ | 4.15:6.6 ± 0.5·10⁻⁴ |
| 10 | 190 ± 0.3 | 32 ± 0.3 | 1.8 ± 0.5·10⁻⁴ | 2.97:3.17 ± 0.5·10⁻⁴ | 3.8:4.1 ± 0.5·10⁻⁴ |
| 11 | 190 ± 0.3 | 32:33 ± 0.3 | 2.57:2.61 ± 0.5·10⁻⁴ | 2.89:2.97 ± 0.5·10⁻⁴ | 1.13:1.34 ± 0.5·10⁻⁴ |
| 12 | 190 ± 0.3 | 33 ± 0.3 | 4.79:4.89 ± 0.5·10⁻⁴ | 3.0 ± 0.5·10⁻⁴ | 0.04 ± 0.5·10⁻⁴ |

Precision in measuring gap and thickness is here higher than in case of in-plane actuators. Nevertheless, specimens exhibit some differences in length, thickness and gap. This layout has two peculiarities. Counter-electrode only partially fills the gap, and the anchor is a structural component with a defined geometry. A crucial aspect was the initial curvature of geometry 9,10 and 11 due to some differences of diffusion of Chromium of the seed-layer among the deposited layers. Models had to include this curvature to fit experimental pull-in voltage. In fact, while microfabrication may induce a residual stress gradient across the beam section in double clamped beam, in microcantilevers stress vanishes since it imposes an initial strain and curvature which moves the free tip. For given initial strain $\varepsilon_0$, curvature $\kappa_0$, accidental thermal effects and axial or flexural preloads, $N_0$ and $M_0$ respectively, stress-strain relations integrated at beam's cross section become:

$$\begin{Bmatrix} N \\ M_z \end{Bmatrix} = \begin{bmatrix} EA & 0 \\ 0 & EJ \end{bmatrix} \left( \begin{Bmatrix} \varepsilon_0 \\ \kappa_0 \end{Bmatrix} - \begin{Bmatrix} \varepsilon_{0T} \\ \kappa_{0T} \end{Bmatrix} \right) + \begin{Bmatrix} N_0 \\ M_0 \end{Bmatrix} \quad (1)$$

where thermal effects are:

$$\varepsilon_{0T} = \alpha T_0(x); \quad \kappa_{0T} = \alpha \int_0^l y T(x,y) dA \quad (2)$$

Symbols mean Young modulus, $E$, beam section, $A$, beam second moment of area, $J$, axial effort, $N$, bending moment, $M$. Temperature distribution may include a constant contribution $T_0$ along beam thickness ($y$ axis), and a distribution $T(x,y)$ variable along beam length ($x$) and thickness ($y$).

Experimental results were similar for the four geometries tested. Those of geometry 10 are depicted in Fig.9. FEM and FEM/BEM approaches converge to a numerical solution which overestimates the actual pull-in voltage. In this case benefits of in-plane layout are absent. Fringing effect is more relevant. To fit experiments it was required to perform a 3D FEM analysis and compute the correction factor for the electromechanical force. It was observed that electric field in 2D models, including separately microbeam length and width, gave easily this number. FEM 3D models did show some problems because of mesh morphing operation applied to a so narrow gap. A sensitivity analysis confirmed that electrostatic pressure is quite uniformly distributed along the length and the width of the microbeam, while only peak values of force strongly depend on thickness and gap values. In fact, an electrical analysis on the undeformed configuration of the microbeam allowed computing a correction factor suitable to find experimental pull-in.

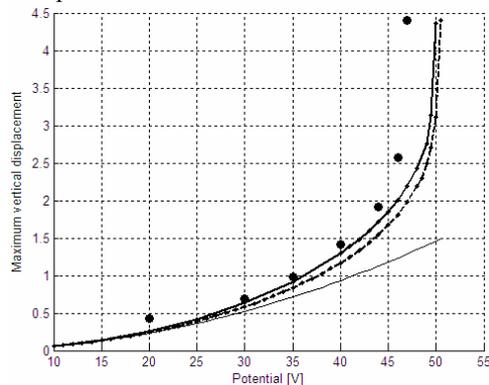

Fig. 9. Example of experimental validation on out-of-plane microcantilever (geometry 10). Experiments (black point) are compared to nonlinear FEM (bold line), nonlinear FEM/BEM (dashed) an geometrically linear solution (grey continuous line).

## IV. DISCUSSION

Model order reduction of nonlinear and second order dynamic microsystems is still a difficult task. No definitive approaches were successfully tested, although some methods demonstrated to be effective in some specific application [1,2,3,16,17]. Nonlinearity is a crucial aspect and microcantilevers exhibit both electromechanical and geometric nonlinearities. Choice is either solving an analytical formulation of the coupled problem, by reducing DoFs involved, or resorting to a numerical sequential solution. In this case mesh morphing inhibits the use of MOR methods [17]. *Ad hoc* linearization was already proposed, in absence of geometric nonlinearity [19], while geometric nonlinear MEMS can be characterized by 2D static models. These allow identifying microsystem stiffness to be used together with damping for dynamic analysis. For in-plane configuration 3D effects of electric field are less relevant, but a very accurate measure of design parameters is required. In case of out-of-plane layout it is just the opposite. Experimental validation demonstrated that force input for 2D models can be calibrated on 3D FEM electrical analyses. Dynamic analyses can be then performed.

## V. CONCLUSION

Two dimensional models are often considered poorly effective to predict static and dynamic behaviour of microbeam RF-MEMS. Microcantilevers with in-plane or out-of-plane bending can be accurately and fairly fast analysed by 2D models, based on sequential non incremental approach implemented in FEM, FEM/BEM or DGA. 2D model tuning can be done by performing a FEM analysis of three dimensional effects of electric field. Geometric nonlinearity is relevant for all the specimens tested. In-plane microcantilevers analysis suffers any inaccuracy in measuring the parameters used as inputs. For out-of-plane microbenders fringing and three dimensional effects of electric field have to be carefully evaluated together with the initial curvature, often present.

Where model order reduction techniques fail because of the double nonlinearity of actuation and geometrical effects, 2D models appear suitable to extract few lumped parameters to perform dynamic analysis. This procedure requires an evaluation of electric field singularities to correct the electrostatic force input.

ACKNOWLEDGMENT

This work was partially funded by the Italian Ministry of University under grant 2005/2005091729, Operative Unit of Udine (p.i. E.Brusa).